\documentstyle[11pt,paspconf,epsf]{article}

\begin{document}

\title{Neutral hydrogen at high redshift: probing structure formation}
\author{J.S.Bagla}
\affil{Institute of Astronomy, Madingley Road, Cambridge CB3 0HA,
        U.K.\\
       E-mail: jasjeet@ast.cam.ac.uk}

\begin{abstract}
Large inhomogeneities in neutral hydrogen in the universe can be
detected at redshifts $z \leq 10$ using the redshifted $21$cm line
emission from atomic hydrogen.  This paper reviews the expected
evolution of neutral hydrogen and presents estimates for future
surveys of HI at $z \approx 3$.  We also discuss the possibility of
detecting neutral hydrogen at higher redshifts.
\end{abstract}

\keywords{Galaxies : formation -- cosmology : theory -- early Universe,
large scale structure of the Universe}

\section{Introduction}

It is generally believed that large scale structures like galaxies and
clusters of galaxies formed from small initial inhomogeneities via
gravitational collapse.  This means that neutral hydrogen was
distributed homogeneously at very high redshifts ($z \gg 20$).  After
recombination, the fraction of neutral hydrogen in molecular form
increases with decreasing redshift (Tegmark et al. 1997).  This leads
to formation of clusters of stars ($M_{cl} \simeq 10^6 M_\odot$,
$z\simeq 20$).  UV radiation from these clusters dissociates H$_2$
molecules (Haiman, Rees and Loeb 1997) preventing formation of
more star clusters via molecular cooling as the Jeans mass increases
by about two orders of magnitude.  

The Inter-Galactic Medium (IGM) is completely ionized at $z\le 5$
(Giallongo et al. 1994) and neutral hydrogen exists only in dense
clumps where the large column density prevents ionizing radiation from
penetrating the inner regions.  

The Universe is reionized between the formation of first star clusters
at $z \simeq 20$ and the highest redshift quasars at $z\simeq 5$.  The
epoch of reionization, and the distribution of sources of ionizing
radiation, leave their imprint on the distribution of neutral and
ionized gases.  The distribution of hydrogen is largely homogeneous
before a significant fraction of the volume is ionized.  In the
intermediate regime, when the Universe is partially ionized, neutral
hydrogen exists in two phases: A warm phase that has been heated
($T_{spin} \gg T_{cmb}$) by radiation from ionizing sources.  This
phase surrounds the ionized regions.  The cold phase of neutral
hydrogen ($T_{spin} \approx T_{cmb}$) can be found in regions far away
from sources of ionizing radiation.  The warm phase can be observed in
redshifted $21$cm emission whereas no such radiation is expected
from the ionized or the cold phases of gas.  This patchiness will
certainly exist at observable angular scales if quasars are the main
source of ionizing radiation (Madau, Meiksin and Rees 1997).  

After reionization of the IGM, neutral hydrogen distribution traces
the distribution of collapsed dark matter halos with circular velocity
$v_c > 50$km~s$^{-1}$ (Thoul and Weinberg 1996).  In this regime, it
is possible to estimate the mass in neutral hydrogen in clumps from 
models of galaxy formation.  Proto-clusters, or large groups
of such halos may be observed in the redshifted $21$cm line emitted by
the neutral hydrogen in these structures.

Thus there are two regimes in which neutral hydrogen can be observed
at high redshifts.  Early epochs, when the Universe is being ionized
and patchiness in the ionization and temperature makes a part of the
universe visible in the redshifted 21cm band, and, late epochs when
neutral hydrogen in self gravitating dense clumps is the main source
of signal.  We will discuss the prospects of observing redshifted
$21$cm line from post-reionization epochs first and then discuss some
aspects of the early, pre-reionization epochs.

\section{Neutral Hydrogen after IGM is Reionized}

After reionization of IGM is complete, neutral hydrogen survives only
in high density, radiatively cooled objects (Weinberg et al. 1996).
Such systems are expected to trace the distribution of galaxies and
can exist only in regions where the density of dark matter is well
above average.  Thus the large scale distribution of neutral hydrogen,
and the distribution of dense halos of dark matter is the same.  

Subramanian and Padmanabhan (1993) computed the expected flux from
proto-clusters at $z \simeq 3$ for some models of structure formation.
They used the Press-Schechter formalism (Press and Schechter, 1975) to
compute the expected number density of proto-clusters in the CDM and
HDM models.  This estimate was a pilot study for the Giant Meter-wave
Radio Telescope (GMRT)\footnote{The GMRT will be able to probe the
  redshifted $21$cm line from three epochs centered at $z=3.34$, $5.1$
  and $8.5$.} presently being constructed in India (Swarup, 1984).  
Modeling was refined in a later paper (Kumar, Padmanabhan and
Subramanian, 1995) where they computed line profiles assuming the
proto-clusters to be spherical perturbations composed of virialized
clumps of smaller masses.  These studies suggest
that it should be possible to detect proto-clusters in the COBE
normalized standard CDM model using the GMRT with $10$ to $20$ hours
of observations.  However, structures that are expected to contribute
strong signal are very rare peaks in the density field and are
expected to occur in about every fifth field of view.

Some authors have studied the distribution of neutral hydrogen at high
redshifts using simulations that include gas dynamics, ionization and
other astrophysical processes.  Most of these studies focus on small
scale variations in the distribution of neutral hydrogen.  [e.g. see
Weinberg et al. (1996)]  However, the synthesized beam, at frequencies
suited for detecting redshifted $21$cm line from $z \ge 3$, for most
telescopes available at present includes a large comoving volume and 
so the details of physical processes operating at small scales can be
ignored to a large extent.  The sensitivity of these telescopes is
also not sufficient to probe small scale structure.  Thus we can
ignore the differences in distribution of baryons and dark matter at
small scales.  Further, as we are interested in the neutral fraction
at a given epoch, we can choose to ignore the physical processes
responsible for its evolution.  This simplifies the problem to a large
extent and we can get meaningful estimates of the signal strength
without a detailed treatment of baryons and astrophysical processes. 

\subsection{Modeling}

Observations of the damped Lyman-$\alpha$ absorption systems (DLAS),
believed to be progenitors of present day galaxies, suggest that a
large fraction of mass in these is in form of HI (Wolfe
et al. 1995).  Observations also show that the spin temperature of gas
in DLAS, the relevant quantity for the $21$cm radiation, is much
higher than the temperature of the background radiation at this
epoch.  Large HI fraction can also be inferred indirectly from the
tentative estimates of star formation history (Connolly et al. 1997).
These estimates suggest that the peak of star formation activity is
around $z \approx 2$, thus a large fraction of the gas may be in the
neutral form at redshifts higher than this.

We can summarize a set of reasonable assumptions that can be used to
estimate the distribution of neutral hydrogen at high redshifts, and
the expected signal in the redshifted $21$cm line.
\begin{itemize}
\item  Neutral hydrogen exists only in highly over dense regions.  Thus
  the distribution of neutral hydrogen is the same as the distribution
  of dense halos.
\item  Neutral hydrogen shares the velocity field of the dark matter
  distribution, and the velocity dispersion of halos in which it
  resides. 
\item  Neutral fraction in these regions is expected to be high at $z
  \ge 2$, i.e. before the star formation activity reached its peak
  level.
\end{itemize}
An additional fact that is of considerable importance is that the
primary beam of most telescopes that have, or will have, the
sensitivity to detect neutral hydrogen at high redshift, contains a
very large comoving volume.  For example, the primary beam for the
GMRT includes a comoving volume of $\approx 10^6$~h$^{-3}$Mpc$^3$.
Therefore, probability of finding a rare peak, like proto-clusters of
mass $\approx 10^{14}M_\odot$ is reasonably high.  Any
implementation of the method outlined above must treat these rare
objects in a fair manner as these will be the first objects to be
detected in redshifted $21$cm emission line.\footnote{This is taken
  into account by using a simulation volume that is comparable to
  the volume enclosed in a primary beam of the GMRT.}

\subsection{Results}

The spatial distribution of neutral hydrogen, when combined with the
velocity field, gives the redshift space distribution.  This can then
be used to determine the expected flux in a fairly straightforward
manner.  If the spin temperature is much greater than the temperature
of the CMBR then the 
spin temperature drops out of the expression for the emitted energy,
which then is proportional to the mass of neutral hydrogen.  The
observed flux can be written in terms of the mass in atomic hydrogen
and the velocity width of this distribution.
\begin{equation}
S_\nu \simeq 220 \mu\hbox{Jy}
\left(\frac{M_{HI}}{10^{13}M_\odot}\right) \left(\frac{1
    \hbox{MHz}}{\Delta\nu_0}\right)
\left(\frac{D_L(z=3.34,h=0.5,\Omega_0=1, \Lambda=0)}{D_L(z)}\right)^2
\end{equation}
Here $M_{HI}$ is the mass in neutral hydrogen in the atomic form,
$\Delta\nu_0$ is the spread in frequency space in the observers frame
and is corresponds to a rest frame velocity width of $v_{disp} \approx
1000$~km s$^{-1}$ at redshift $z=3.34$ and $D_L$ is the luminosity
distance.  For reference, we have used the luminosity distance for an
Einstein-de~Sitter Universe with $h=.5$ at $z=3.34$.  We can write for
$M_{HI}$, 
\begin{equation}
M_{HI} = M_{total} \; f_{neutral} \; \frac{\Omega_b}{\Omega_0} =
0.025 \; M_{total} \; \left(\frac{f_{neutral}}{0.5}\right) \;
\left(\frac{\Omega_b}{0.05}\right) \; \left(\frac{\Omega_0}{1}\right) 
\end{equation} 

\begin{figure}
\plotone{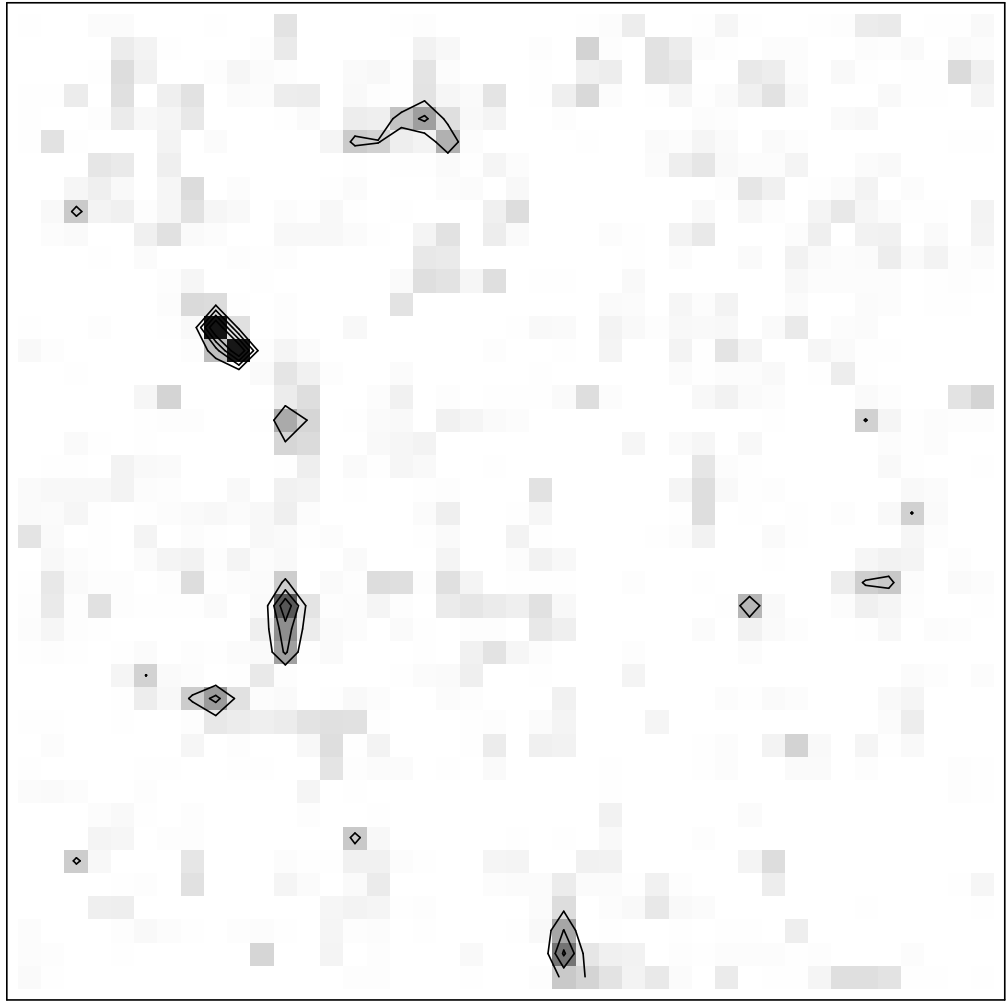}
\caption{This figure shows a sample radio map ($z=3.34$ from a
  simulation of the standard CDM model.  Angular size of each pixel is
  $3.2'$, corresponding to about $3$h$^{-1}$Mpc (comoving), and the
  bandwidth of this image is $125$kHz.  The contour levels correspond
  to a signal of $80$, $60$, $40$ and $20$~$\mu$Jy. A density
  threshold of $4\bar\varrho$ was used in constructing this map.}
\end{figure}

We can compare the expression for flux with the sensitivity of the
GMRT.\footnote{These numbers are for the central square which has $12$
  antennas spread in a region of $1$sq~km.} 
\begin{eqnarray}
\hbox{rms noise} &=& 44 \mu Jy \left(\frac{T_s}{110K}\right) \left(\frac{1
MHz}{\Delta \nu} \right)^{1/2} \left( \frac{100 hrs}{\tau} \right)^{1/2}
\nonumber \\
&=& 100 \mu Jy \left(\frac{T_s}{250K}\right) \left(\frac{1
MHz}{\Delta \nu} \right)^{1/2} \left( \frac{100 hrs}{\tau} \right)^{1/2}
\end{eqnarray}
Thus it should be possible to detect a proto-condensate of mass
$M_{HI} = 10^{13}M_\odot$ or greater with about $100$ hours of
observations.  In the context of existing constraints (Wieringa, de
Bruyn and Katgert 1992), GMRT will be able to probe mass scales that
are about an order of magnitude smaller than those accessible to
previous surveys at redshift $z \approx 3$.  The square kilometer
array being planned (Braun 1996) is expected to have the sensitivity
to detect clumps of neutral hydrogen above $10^9 M_\odot$.

Bagla, Nath and Padmanabhan (1997) studied some models
of structure formation and generated radio maps from N-Body
simulations of dark matter using the assumptions outlined above.  The
highest peaks in most models studied there have a flux around $150
\mu$Jy for the redshift window $z=3.34$. 

Figure 1 shows a radio map constructed from a simulation of the
standard CDM model.  The power spectrum was normalized to match the
abundance of clusters at the present epoch, i.e. $\sigma_8=0.6$.  The
map shows angular distribution of flux from a region of width
$0.125$kHz at $z=3.34$.  Pixels are $3.2'$ wide, corresponding to
about $3$h$^{-1}$Mpc (comoving).  The contour levels correspond to a
signal of $80$, $60$, $40$ and $20$~$\mu$Jy. A density threshold of
$4\bar\varrho$ was used in constructing this map.  The highest peak
encountered in this simulation had a flux of $115\mu$Jy and a FWHM of
about $1$MHz.  The contour for half signal enclosed three pixels.  We
used a neutral fraction $f_N = 0.5$.

If the assumption of complete reionization of the IGM is valid at
$z\approx 5$ then the expected signal, for the same models is about
$200\mu$Jy, which can also be observed with $100-200$ hours of
observations. 

In general, the expected signal for a fixed neutral fraction depends
mainly on the amplitude of fluctuations at the cluster scale.  The
angular size of typical proto-condensates depends on the slope of the
power spectrum at cluster scales.  Power spectra with more power at
smaller scales give rise to more concentrated proto-condensates.

Thus a direct detection of proto-condensates, at $z \simeq 3$ and
$z\simeq 5$ should be possible with observations of about $100-200$hrs
if we search for a spectral line in emission.  Another possible method
is to look for variations in rms flux received from different
directions.  This method is quicker, but may not be practical if
confusion due to fainter sources dominates at these angular scales.
Using high resolution imaging to reduce the contribution of confusion
may make this method feasible for detecting neutral hydrogen.  This
may be possible, in the case of the GMRT, if the full array is used
for collecting data.  Then the longer baselines can be used to
remove discrete sources.  This method will also reduce the integration
time required by a significant amount as all $30$ antennas can be used
instead of the $12$ in the central square.  As an aside, we would like
to point out that it will be possible to detect objects in the
continuum emission down to a few tens of $\mu$Jy, making any observed
field the most well studied one at low frequencies.  These sources
will include faint AGNs and star forming galaxies.  Multi-frequency
observations of the same field will provide considerable wealth of
information about these sources, making it a worthwhile project for
more than one reason.

Recent observations of Lyman break galaxies at $z \approx 3$ have
revealed large spikes in their redshift space distribution (Steidel et
al. 1998).  The estimated mass contained in these spikes is comparable
to $10^{15} M_\odot$ and hence these are natural targets for searches
of neutral hydrogen at $z \approx 3$.  These observations also
suggest that the probability of finding such spikes at high redshifts
is large.

\section{Neutral Hydrogen before reionization of IGM is complete}

Hydrogen in the Universe exists in three phases at epochs before
reionization is complete.  The ionized phase in and
around dense clumps where sources of ionizing radiation are present:
young stars, and/or, quasars.  The warm phase that is expected to
envelope the ionized phase, these regions have neutral hydrogen with
$T_{spin} \gg T_{cmb}$.  Rest of the Universe contains cold gas with
$T_{spin} \approx T_{cmb}$.  It is possible, in principle, to observe
the warm phase in redshifted $21$cm line (Madau, Meiksin and Rees
1997).  These will appear as patchy shells around (invisible) sources
of ionizing radiation.  The typical separation of shell centers can be
used as an indicator of the number density of ionizing sources.  The
separation will be large if the sources are rare objects like quasars.
However, it will be small if most of the ionizing radiation comes from
stars. It is expected that these early structures will cluster
strongly (Bagla 1997) and hence the typical separation between shell
centers will be many times larger than $n^{-1/3}$ where $n$ is the
number density of sources.

Observability of patchiness at high redshifts require observations at
very low frequencies, e.g., $160$MHz for $z \approx 8$ and $70$MHz for
$z \approx 20$.  At these low frequencies, and at the very low flux
levels, the ionosphere and galactic emission are the main sources of
noise.  In the near future, it will be possible to attempt
observations at $150$MHz using the GMRT.

\section{Summary}

Detection of proto-clusters at $z \approx 3$ should be possible in
near future.  This requires integration of about $100$hrs with the
existing and upcoming telescopes, the numbers presented here were
computed specifically for the GMRT.

If interferometric observations can be carried out at $233$MHz for the
low flux levels expected from proto-clusters then it may be possible
to detect proto-clusters at $z \approx 5$ as well.

Patchiness in the pre-reionization stage may be observed using the
$21$cm tomography (Madau, Meiksin and Rees 1997) at $z \approx 8$ if
low flux levels can be observed at $150$MHz.  This may be possible, if
at all, only during solar minimum when the ionosphere is relatively
stable. 

Any search for neutral hydrogen at high redshifts will have spin-offs
like observations of star burst galaxies and faint AGNs to very low
flux levels.

\acknowledgements

I acknowledge the support of PPARC fellowship at the Institute of
Astronomy, Cambridge.  I thank T.Padmanabhan, Jayaram Chenglur and
Martin Rees for useful discussions.

\end{document}